\documentclass[%
superscriptaddress,
preprint,
 amsmath,amssymb,
 aps,
]{revtex4-1}

\usepackage{graphicx}% Include figure files
\usepackage{dcolumn}% Align table columns on decimal point
\usepackage{bm}% bold math
\usepackage{color}%color

\begin{document}

%\preprint{APS/123-QED}

\title{Gate tunable optical absorption and band structure of twisted bilayer graphene}% Force line breaks with \\

\author{Kwangnam Yu}
\affiliation{Department of Physics, University of Seoul, Seoul 130-743, Korea}

\author{NGUYEN Van Luan}
\affiliation{Center for Integrated Nanostructure Physics, Institute for Basic Science (IBS), Sungkyunkwan University, Suwon 16419, Korea}

\author{Tae Soo Kim}
\affiliation{Center for Integrated Nanostructure Physics, Institute for Basic Science (IBS), Sungkyunkwan University, Suwon 16419, Korea}

\author{Jiwon Jeon}
\affiliation{Department of Physics, University of Seoul, Seoul 130-743, Korea}

\author{Jiho Kim}
\affiliation{Department of Physics, University of Seoul, Seoul 130-743, Korea}

\author{Pilkyung Moon}
\email{Corresponding author 1, e-mail : pilkyung.moon@nyu.edu}
\affiliation{New York University and NYU-ECNU Institute of Physics at NYU Shanghai, Shanghai 200122, China}
\affiliation{Department of Physics, New York University, New York, NY 10003, USA}

\author{Young Hee Lee}
\email{Corresponding author 2, e-mail : leeyoung@skku.edu}
\affiliation{Center for Integrated Nanostructure Physics, Institute for Basic Science (IBS), Sungkyunkwan University, Suwon 16419, Korea}

\author{E. J. Choi}
\email{Corresponding author 3, e-mail : echoi@uos.ac.kr}
\affiliation{Department of Physics, University of Seoul, Seoul 130-743, Korea}

\date{\today}% It is always \today, today,
             %  but any date may be explicitly specified

\begin{abstract}
We report the infrared transmission measurement on electrically gated twisted bilayer graphene.
The optical absorption spectrum clearly manifests the dramatic changes such as the splitting of inter-linear-band absorption step, 
the shift of inter-van Hove singularity transition peak, and the emergence of very strong intra-valence (intra-conduction) band transition.
These anomalous optical behaviors demonstrate consistently the non-rigid band structure modification created by the ion-gel gating 
through the layer-dependent Coulomb screening.
We propose that this screening-driven band modification is an universal phenomenon that persists to other bilayer crystals in general,
establishing the electrical gating as a versatile technique to engineer the band structures and to create new types of optical absorptions 
that can be exploited in electro-optical device application.
\end{abstract}

% PACS, the Physics and Astronomy
                             % Classification Scheme.
%\keywords{Suggested keywords}%Use showkeys class option if keyword
                              %display desired
\maketitle

%\tableofcontents

{\it Introduction.}
The chemical potential $\mu$ of two-dimensional materials composed of one or a few atomic layers exhibits a marked shift 
by electrical gating due to their ultra-thin sample thickness \cite{novoselov2004electric,novoselov2005two,oostinga2008gate,furchi2014mechanisms,costanzo2016gate},
thus manifesting numerous novel optical properties under external bias.
In monolayer graphene, for example, the universal optical conductivity under the linear band regime $\sigma_{\mathrm{mono}} = e^2 / 4\hbar$ \cite{nair2008fine} 
can be tailored at a certain photon energy $E$ by controlling $\mu$ \cite{li2008dirac,mak2008measurement, jiang2018gate}, allowing for a graphene-based optical modulator \cite{liu2011graphene}.
The electrical tuning of optical absorption is not limited to monolayer graphene but was also observed in other two-dimensional materials such as 
Bernal-stacked bilayer graphene and black phosphorus \cite{zhang2009direct, li2009band, chernikov2015electrical, whitney2016field, ju2017tunable, yao2017optically}.

\begin{figure*}[ht]
\centering
\includegraphics[width=1.0\linewidth]{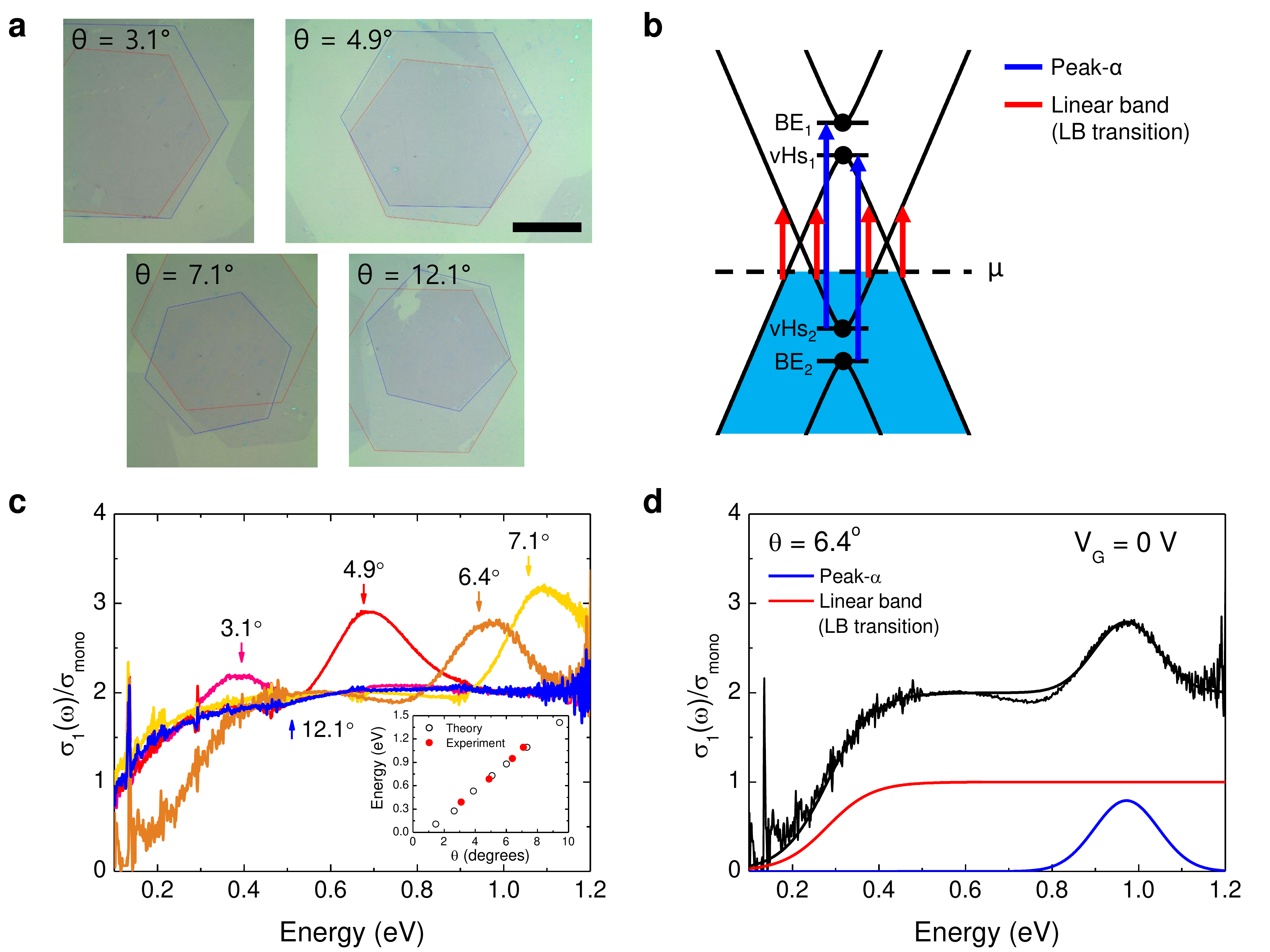}
\caption{
Optical absorption of ungated tBG. 
(a) Optical images of tBG samples with different twisted angle $\theta$. Scale bar is 20~$\mu$m.
(b) Electronic band diagram of tBG.
BE and vHs stand for the band edge of the second band and the saddle point van Hove singularity, respectively.
Two kinds of optical transitions are activated as shown by the red and blue arrows.
(c) Optical conductivity $\sigma_{1}(\omega)$ of the five tBG samples.  
The Peak-$\alpha$ shifts to higher energy as $\theta$ increases.  
The sharp peak at E = 0.13~eV is an artifact due to optical phonon absorption of SiO$_{2}$.
Inset compares the measured peak position with the theoretical prediction \cite{moon2013optical}. 
(d)
The $\sigma_{1}(\omega)$ can be fit in terms of
the LB-transition (red curve) and Peak-$\alpha$ (blue curve). 
}
\label{fig:1}
\end{figure*}

A twisted bilayer graphene (tBG), two sheets of graphene stacked with a twist angle $\theta$, 
has attracted a great deal of attention due to the fascinating physics such as the moir\'e superlattice, Hofstadter butterfly, 
and the emergence of two-dimensional superconductivity \cite{ohta2012evidence,bistritzer2011moire,moon2012energy,cao2018unconventional}.
When the Fermi energy lies at the charge neutrality point ($\mu = 0$),
the low-energy optical spectrum of tBG is characterized by the linear-band absorption from the two graphene layers 
[indicated by the red arrows in figure~\ref{fig:1}(b), LB hereafter], yielding $2\sigma_{\mathrm{mono}}$.
At higher energies, however, the interlayer interaction hybridizes the LBs of the two graphene layers
and yields a band anticrossing as illustrated in figure~\ref{fig:1}(b) \cite{dos2007graphene,mele2010commensuration,dos2012continuum,tabert2013optical,moon2013optical}.
Although the transition between the saddle point van Hove singularities (vHs), i.e., $\mathrm{vHs}_2 \rightarrow \mathrm{vHs}_1$, 
is exactly forbidden by the lattice symmetry \cite{moon2013optical}, the transition between vHs and the band edge (BE) of the second band exhibits 
a prominent absorption peak coming from the large joint density of states 
[blue arrows in figure~\ref{fig:1}(b), Peak-$\alpha$ hereafter] \cite{tabert2013optical,moon2013optical,havener2014van,kim2016chiral}.

It is expected that the optical absorption spectrum of tBG under electrical gating will exhibit rich physics compared to that of monolayer graphene.
For example, if $\mu$ reaches vHs$_{2}$ (vHs$_{1}$), either of the interband transitions Peak-$\alpha$ will vanish 
and a new intra-valence (intra-conduction) band transition,	such as $\mathrm{BE}_2 \rightarrow \mathrm{vHs}_2$ ($\mathrm{vHs}_1 \rightarrow \mathrm{BE}_1$),
will emerge due to the depletion of electron (hole).
However, such dramatic changes have never been experimentally observed yet.
Here we report the first infrared transmission measurement of gated tBG,
and show the optical absorption spectrum with a varying $\mu$ over a wide range.
Our result demonstrates that the electrical gating leads to a significant modification of band structures, 
in addition to the $\mu$-shift, by creating an interlayer potential asymmetry between the two graphene layers.
We further elucidate the full non-rigid band modification schemes and discuss its implication on optical device application.

{\it Experiment.}
To make a tBG sample, we first prepared a single-domain monolayer graphene by the chemical vapor deposition (CVD) on a single crystal Cu substrate \cite{nguyen2015seamless}.
Then a pre-grown second graphene sheet was transferred on top of the first one by the bubbling and alignment technique \cite{nguyen2016wafer}.
The tBG sheet was transferred on SiO$_{2}$/Si substrate for further optical measurement.
We chose lightly p-doped Si substrate (resistivity $\sim 10~\Omega$cm) which is IR transparent, which allowed us conduct the transmission measurements \cite{horng2011drude}.
To perform the ion-gel gating on the tBG, the mixture of [EMIM][TFSI] ionic liquid, PS-PMMA-PS triblock copolymer, 
and ethyl acetate solvent (weight ratio $= 0.1:0.9:9$) was prepared and spin-coated on the sample \cite{kim2010high}.
Optical transmittance was measured using a microscopic FTIR (Bruker, Hyperion 2000) for an infrared frequency range on five samples with different $\theta$'s. 
The optical conductivity $\sigma_{1}(\omega)$ of tBG was extracted by fitting the raw data using the multilayer Kramers-Kronig analysis program, RefFit \cite{kuzmenko2005kramers}.

{\it Results.}
In figure~\ref{fig:1}(c), we show the optical conductivity $\sigma_{1}(\omega)$ of tBG with various rotation angles (see  supplemental figure S1 for the raw data).
The $\sigma_{1}(\omega)$ spectrum shows two distinct interband transitions described in figure~\ref{fig:1}(b):
(i) the frequency-independent conductivity $2\sigma_{\mathrm{mono}}$ which comes from the LB-transition (red arrows), and 
(ii) the prominent absorption Peak-$\alpha$ which comes from the transitions vHs$_{2}$ $\rightarrow$ BE$_{1}$ and BE$_{2}$ $\rightarrow$ vHs$_{1}$ \cite{tabert2013optical,moon2013optical}. 
Peak-$\alpha$ blueshifts as $\theta$ increases [figure~\ref{fig:1}(c) and inset], since the two Dirac cones in figure~\ref{fig:1}(b) move apart, 
increasing the energy of the vHs \cite{moon2013optical}.
Figure~\ref{fig:1}(d) shows the fitting of the conductivity with LB and Peak-$\alpha$ using a model function
\begin{equation}
	\sigma_{1}(\omega,\mu) = 2\sigma_{\rm{LB}}(\omega,\mu) + \sigma_{\alpha}(\omega),
	\label{eq:1}
\end{equation}
Here the factor 2 comes from the layer degeneracy, and the LB-transition conductivity for each layer is given by
\begin{equation}
	\sigma_{\rm{LB}}(\omega,\mu) = \sigma_{\mathrm{mono}}
	\times \left[
	\tanh\left(\frac{\hbar\omega+2\mu}{4k_{\rm{B}}T}\right)+\tanh\left(\frac{\hbar\omega-2\mu}{4k_{\rm{B}}T}\right)
	\right] / 2,
	\label{eq:2}
\end{equation}
where $\mu$(= 0.28~eV) and $T$(= 300~K) are the chemical potential and temperature, respectively \cite{mak2008measurement}.
For Peak-$\alpha$, we use the standard Gaussian function
\begin{equation}
	\sigma_{\alpha}(\omega)=\frac{S_{\alpha}}{\sqrt{2\pi\Gamma_{\alpha}^2}}\exp\left[-\frac{(\hbar\omega-E_{\alpha})^2}{2\Gamma_{\alpha}^2}\right],
	\label{eq:3}
\end{equation}
where $S_{\alpha}$, $E_{\alpha}$, $\Gamma_{\alpha}$ stand for the intensity, energy and broadening of the peak, respectively.
The model function fits very well the measured conductivity except for a minor discrepancy at 0.76~eV,
where the latter discrepancy is due to the deactivation of the LB-transition in the small region of K-space 
where the hybridization gap is opened, as theoretically predicted \cite{moon2013optical}.

\begin{figure*}[ht]
\centering
\includegraphics[width=0.5\linewidth]{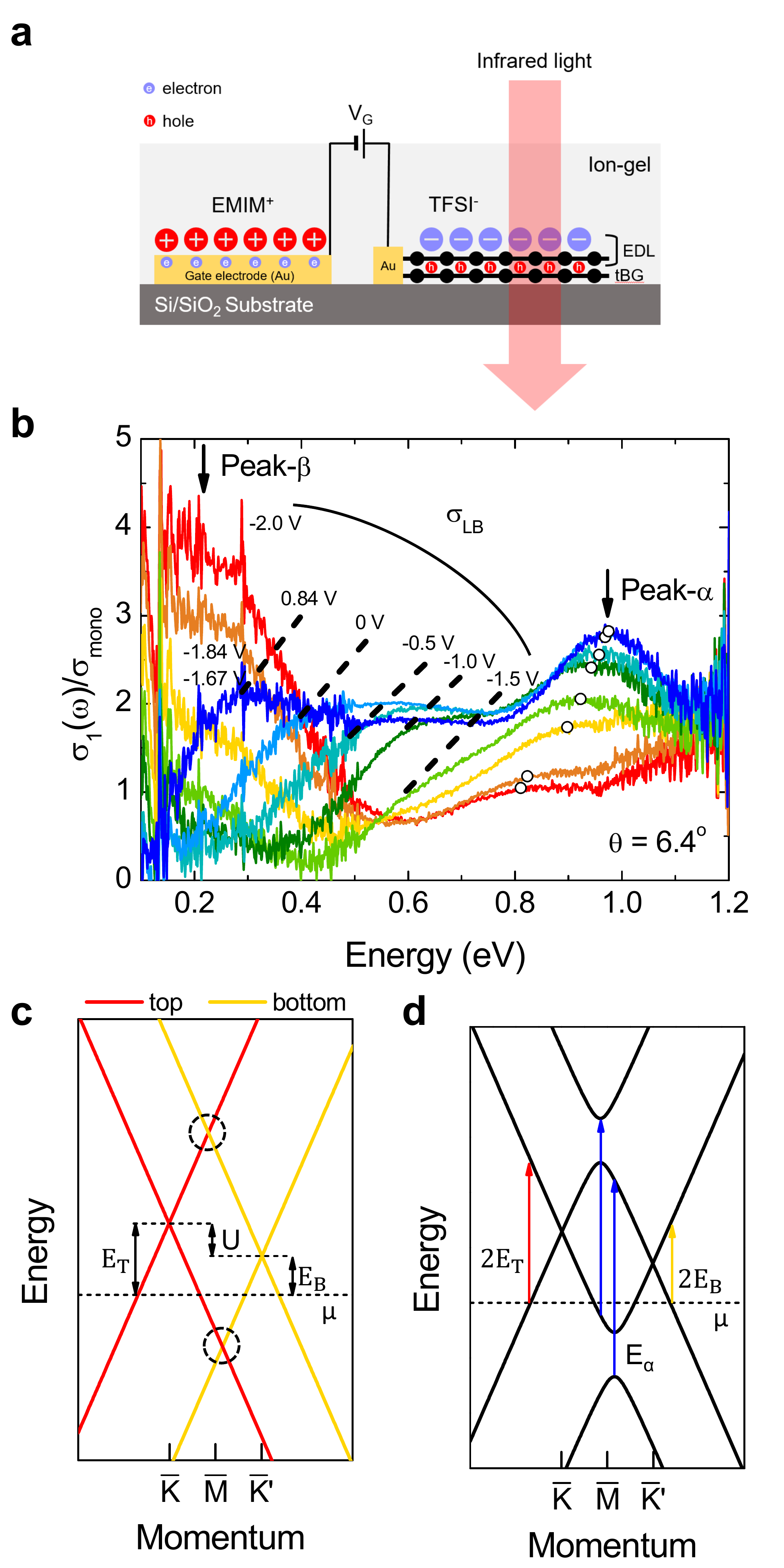}
\caption{
Optical absorption of Ion-gel gated tBG.
(a) Schematic view of the ion-gel gating circuit and the infrared transmission measurement.
(b) Optical conductivity of gated tBG ($\theta=6.4^{\rm{o}}$) for various gate voltage $V_{\rm{G}}$. 
The gate-driven changes of the LB-transition, Peak-$\alpha$, and Peak-$\beta$ are observed and discussed in detail in the text. 
(c) the band structure of tBG under gating. 
The top band and bottom band shift by $E_{\rm{T}}$ and $E_{\rm{B}}$ respectively. 
$U$ is their difference $U=E_{\rm{T}}-E_{\rm{B}}$.
Here the gap opening is omitted for clarity.
(d) Optical transitions change in the gated tBG compared with the un-gated one as result of the band shifts.
}
\label{fig:2}
\end{figure*}

\begin{figure*}[ht]
\centering
\includegraphics[width=0.7\linewidth]{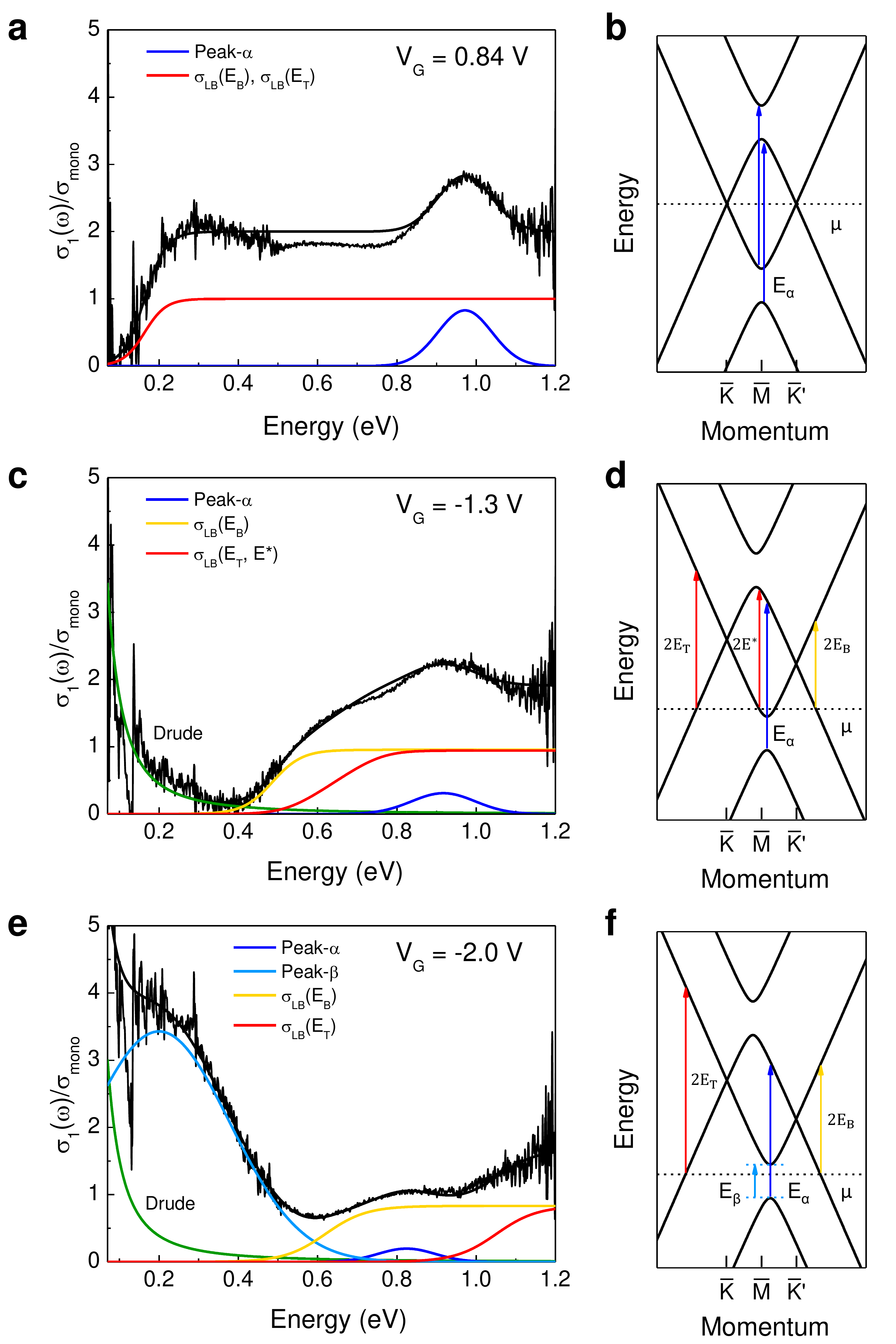}
\caption{
Gate-driven evolution of the optical conductivity (left column) and the band structure (right column).
(a), (c), (e) Fit of optical conductivity for $V_{\rm{G}}=0.84$~V, -1.3~V, and -2.0~V. 
(b), (d), (f) Non-rigid band evolution and optical transitions for (a), (c), and (e).
As $V_{\rm{G}}$ is increased, $U$ becomes stronger and as result Peak-$\alpha$ transition energy decreases.
In (f), the new $\beta$-transition (sky-blue) is activated.
}
\label{fig:3}
\end{figure*}

\begin{figure*}[ht]
\centering
\includegraphics[width=0.55\linewidth]{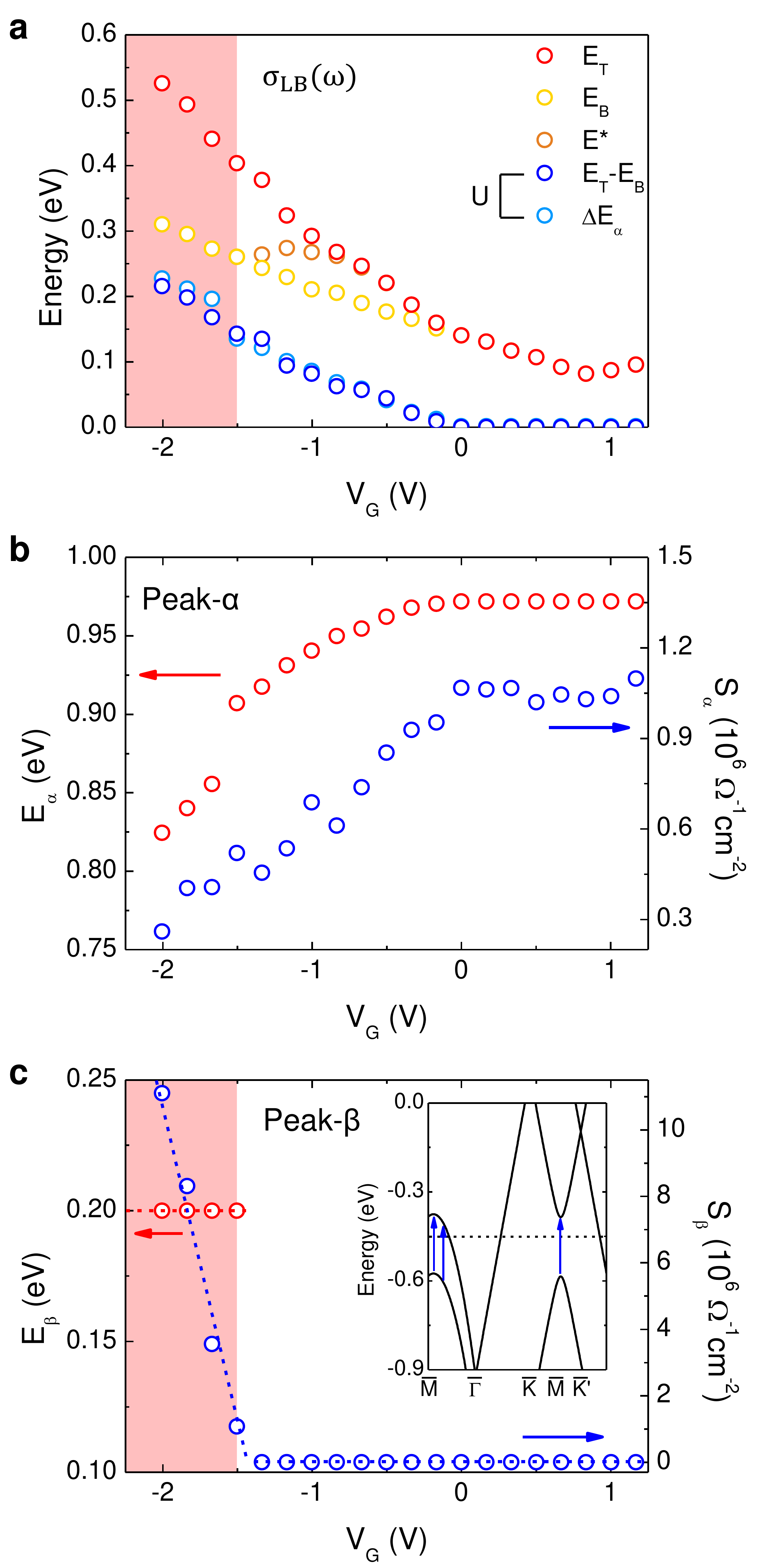}
\caption{
Fitting result of the peak energy and intensity
(a) LB-transition energy $E_{\rm{T}}$ (and $E^{*}$) for the top graphene, and $E_{\rm{B}}$ for bottom graphene.
$U$ is calculated from the LB-transition and, independently, from the Peak-$\alpha$ shift $U=E_{\alpha}(0)-E_{\alpha}(V_{\rm{G}})$. 
(b) Peak energy $E_{\alpha}$ and strength $S_{\alpha}$ of Peak-$\alpha$. 
(c) $E_{\beta}$ and $S_{\beta}$ of Peak-$\beta$.
Inset shows the band structure along $\bar{M} \rightarrow \bar{\Gamma}$ and $\bar{M} \rightarrow \bar{K}$, 
where the arrows emphasizes the optical criticality of Peak-$\beta$. 
}
\label{fig:4}
\end{figure*}

We prepared an ion-gel gating circuit on the $\theta$ = $6.4^{\rm{o}}$ sample as schematically shown in figure~\ref{fig:2}(a), 
and measured the optical transmittance by varying the bias voltage $V_{\rm{G}}$ over $-2~\mathrm{V} < V_{\mathrm{G}} < 2~\mathrm{V}$.
The reference charge neutral point $V_{\rm{CNP}}$ of this tBG sample is 0.84~V as we will show later in figure~\ref{fig:4}(a).
Figure~\ref{fig:2}(b) shows the $V_{\rm{G}}$-driven change of $\sigma_{1}(\omega)$ for the hole-doping regime, $-2~\mathrm{V} < V_{\mathrm{G}} < V_{\mathrm{CNP}}$.
In a rigid-band picture, the absorption edge of LB-transition $\sigma_{\mathrm{LB}}$ will remain sharp, 
and both the energy and intensity of Peak-$\alpha$ will remain unchanged with $V_{\rm{G}}$ until $\mu$ reaches vHs$_{2}$.
The absorption profile in figure~\ref{fig:2}(b), however, shows that
(i) the absorption edge of $\sigma_{\rm{LB}}$ shows a considerable broadening and shifts to higher energy, and 
(ii) Peak-$\alpha$ shifts to lower energy while its intensity is reduced markedly.
Thus, our observations provide clear evidence that the band structure of tBG varies with gating.
Such a drastic change of the band structure, i.e., a non-rigid band modification, mainly arises from the charge screening by the graphene layer.
The two graphene layers have asymmetric charge density distribution, since the electric field from the ion-gel is screened by the charge at the top graphene layer.
Thus, the bands of the top and bottom layers shift by a different amount, $E_{\rm{T}}$ and $E_{\rm{B}}$ respectively,
leading to an asymmetric potential profile as sketched in figure~\ref{fig:2}(c).
Figure~\ref{fig:2}(d) shows the effect of the asymmetric band shift on the optical transitions. 
Firstly the absorption edge of the LB-transition splits into $2E_{\rm{T}}$ and $2E_{\rm{B}}$ for the two graphene layers
rather than being degenerated, demonstrating that the band shift difference $U$ can be determined once the LB-transition energies are measured from $U=E_{\rm{T}}-E_{\rm{B}}$.
Secondly, the hybridization gap in the conduction band and valence band shift from their un-gated position $\bar{M}$ to opposite directions, $\bar{K}$ and $\bar{K}'$, respectively. 
As result, Peak-$\alpha$ energy is reduced. 
Also, since the optical transition can no longer simultaneously involve vHs and BE, the intensity of Peak-$\alpha$ is decreased.
When $\mu$ passes vHs$_{2}$ by strong gating, a new absorption peak (Peak-$\beta$ hereafter) emerges due to the intra-valence band transition as shown in figure~\ref{fig:3}(f).
To reflect these non-rigid band features, we fit the absorption spectrum to
\begin{equation}
	\sigma_{1}(\omega) = \sum_{i=\mathrm{T,B}} \sigma_{\mathrm{LB}}(\omega,E_i)
	+ \sum_{j=\alpha,\beta} \sigma_{j}(\omega) + \sigma_{\rm{D}}(\omega),
	\label{eq:4}
\end{equation}
The first term in Eq.~(\ref{eq:4}) is summed over the top (T) and bottom (B) layers, and represents the LB-transition from two shifted Dirac cones.
The second term refers to the vHs-BE transitions Peak-$\alpha$ and Peak-$\beta$,
in the form of the Gaussian function with the intensity $S_\beta$, energy $E_\beta$, and broadening $\Gamma_\beta$ of the peak, 
and the last term $\sigma_{\rm{D}}(\omega)$ represents the Drude (intra-band) conductivity of Dirac carrier \cite{horng2011drude,lee2011optical,yu2016infrared}.

Figure~\ref{fig:3} shows the fitting result for three representative $V_{\rm{G}}$'s corresponding to CNP (= 0.84~V), intermediate (-1.3~V), and strong (-2~V) gating.
The fits are in fair agreement  with the measured optical absorption spectrum as shown in figure~\ref{fig:3}(a), (c), (e), and reveals the relevant parameters,
$E_\mathrm{T/B}$, $S_{\alpha/\beta}$, $E_{\alpha/\beta}$, and $\Gamma_{\alpha/\beta}$.
Figure~\ref{fig:3}(b), (d), (f) show the band structure for each $V_{\rm{G}}$'s that reproduces the optical transition energies as fitted.

Figure~\ref{fig:4} displays the fitting parameters as function of  $V_{\rm{G}}$. 
Figure~\ref{fig:4}(a) shows that the band shifts $E_{\rm{T}}$ and $E_{\rm{B}}$ increase along with gating, consistent with the band evolution in figure~\ref{fig:3}(b)-(f). 
Here we determine the interlayer potential difference $U$ from $E_{\rm{T}}$ and $E_{\rm{B}}$ by calculating $U = E_{\rm{T}}-E_{\rm{B}}$. 
We also show $E^{*}$ which refers to the lower-bound of the LB-transition of the top band at $\bar{M}$ as depicted in figure~\ref{fig:3}(d). 
$E^{*}$ is smaller than $E_{\rm{T}}$ due to the gap opening.
By refining $\sigma_{\rm{LB}}$ in the fit, we found the lower-bound of the LB-transition, $2E^{*}$, 
for a gating range of $V_{\rm{G}}<-1~\rm{V}$ (see supplemental figure S2 for details).
At weak gating, $E^{*}$ is close to the value of $E_{\rm{T}}$ but it becomes closer to the value of $E_{\rm{B}}$ as $\mu$ reaches vHs$_{2}$.
Thus, $E^{*}$ and $E_{\rm{B}}$ indicate that $\mu$ reaches vHs$_2$ at $V_{\rm{G}} = -1.5~V$.
Figure~\ref{fig:4}(b) shows the $V_{\rm{G}}$-dependence of the energy and intensity of Peak-$\alpha$. 
The peak energy $E_{\alpha}$ decreases with $V_{\rm{G}}$. Figure~\ref{fig:2}(c) and (d) show that the amount of the gate-driven change of $E_{\alpha}$,
$\Delta E_{\alpha}$ [$\equiv E_{\alpha}(0)-E_{\alpha}(V_{\rm{G}})$] is approximately equals to $U$.
Our calculation of $U$  (see supplemental figure S3 for details) is compared with the $U$ obtained from $\sigma_{\mathrm{LB}}$ in figure~\ref{fig:4}(a).  
The two $U$'s calculated from independent optical transitions $\sigma_{\mathrm{LB}}$ and Peak-$\alpha$ are consistent with each other for the whole range of $V_{\rm{G}}$, 
demonstrating that the non-rigid band picture can describe the physical properties of the gated tBG very well.
The intensity of Peak-$\alpha$ ($S_{\alpha}$) is proportional to the optical transition matrix element (${M}_{if}$) and the density of the initial $\rho_{i}$ and final states $\rho_{f}$.
At $V_{\rm{G}} = V_{\rm{CNP}}$, $S_{\alpha}$ is very large since the vHs and BE align in the Brillouin zone giving a very large $\rho_{i} \rho_{f}$.
However, $S_{\alpha}$ decreases as we apply the gate bias since (i) the vHs and BE no longer align due to $U \ne 0$, 
and (ii) one of the two $\alpha$-transition channels becomes silent when $\mu$ enters the gap [figure~\ref{fig:3}(b), (d), (f), and \ref{fig:4}(b)].
Figure~\ref{fig:4}(c) shows the $V_\mathrm{G}$-dependence of the intra-valance band transition, Peak-$\beta$.
Peak-$\beta$ is silent until $\mu$ reaches vHs$_{2}$, i.e., $V_{\mathrm{G}}$ reaches -1.5~V.
Once the transition is activated, however, it gives a strong absorption peak at the energy corresponding to the magnitude of band opening,
i.e., the energy difference between vHs$_{2}$ and BE$_{2}$.
Theoretical investigation predicts that the amount of band opening equals $2u_{0}$, where
\begin{equation}
	u_0 = \frac{1}{\sqrt{S\tilde{S}}} \int T(\textbf{r}+ d_z \textbf{e}_z) e^{-i \textbf{K}\cdot \textbf{r}} d\textbf{r},
	\label{eq:5}
\end{equation}
is the measure of the interlayer interaction strength \cite{dos2007graphene,tabert2013optical,moon2013optical}.
Here, $S$ and $\tilde{S}$ are the area of each graphene layer, $T(\textbf{R})$ is the transfer integral between the atoms at a relative vector of $\textbf{R}$,
$d_{z}$ is the distance between the two layers, $\textbf{K}$ is the Dirac point of graphene, and the integral in $\textbf{r}$ is taken over the total two-dimensional space.
The band structures of low-angle tBGs ($\theta \le 10^{\circ}$) are very well described by this single parameter, $u_{0}$.
The energy of Peak-$\beta$, $E_{\beta} = 0.2~{\rm{eV}} (= 2u_{0}$) measured in this work, provides an important information about its value of 0.1~eV,
which is consistent with the theoretical expectation of $u_{0} \sim 0.110$~eV.
In addition, figure~\ref{fig:4}(c) shows that the intensity $S_\beta$ grows rapidly with further gating.
This is because the two bands associated with this transition are almost parallel along $\bar{M}-\bar{\Gamma}$ [see the inset of figure~\ref{fig:4}(c)] \cite{moon2014optical},
$\nabla_{\bf{K}}E_{f}\cong\nabla_{\bf{K}}E_{i}$, which give a prominent peak intensity 
$S_{\beta} \sim \int_{FS} \frac{dS}{|\nabla_{\bf{K}}E_{f}-\nabla_{\bf{K}}E_{i}|}$ (FS = Fermi surface). 
From the band structures, we expect that Peak-$\beta$ will grow further for deeper gating $V_{\rm{G}}<-2\,\mathrm{V}$. 
Theoretically $t$ is predicted to be independent on carrier density \cite{moon2014optical}, which is supported by our $E_{\beta}$ being constant with $V_{\rm{G}}$.

{\it Discussion.}
We investigated the optical absorption spectrum of electrically gated tBG.
We showed that two different kinds of interband optical transitions take place in tBG, namely, 
the transition between the linear bands ($\sigma_{\rm{LB}}$) and that between vHs and the band edge (Peak-$\alpha$).
Their behaviors with varying applied bias show the clear evidence of the non-rigid band evolution.
Specifically, the absorption edge of $\sigma_{\rm{LB}}$ is split into two edges with different energies,
indicating that the Dirac cones of the top and bottom layers are shifted by different amount of energy.
In addition, both the energy and intensity of Peak-$\alpha$ show marked changes with the gate bias,
since the interlayer potential asymmetry breaks the alignment of the vHs and the band edge associated with the hybridization gap changing the band structures from direct to indirect-like.
The amounts of interlayer potential asymmetry $U$ extracted from the two independent phenomena, $\sigma_{\rm{LB}}$ and Peak-$\alpha$, were consistent with each other.
Besides, we found that a unique intra-valence (intra-conduction) band transition (Peak-$\beta$) is activated with further gating.
The intensity of Peak-$\beta$ can grow much larger than that of Peak-$\alpha$,
and the energy of Peak-$\beta$ is, unlike to Peak-$\alpha$, less sensitive to the gate bias after it is activated.
Moreover, the energy of Peak-$\beta$ provides an important information on the interlayer interaction strength,
which is essential in revealing the full band structures of tBG.

The gate tunable optical absorption has significant impact on electro-optical application.
For Peak-$\beta$, tBG can either transmit ($S_{\beta}=0$) or strongly absorb (large $S_{\beta}$) the infrared light at $E_{\beta}= 0.2$~eV 
by the $V_{\rm{G}}$-control, which can be used for tunable modulator or filter. 
As for Peak-$\alpha$, this peak absorbs visible light (red, green, blue) for tBG samples in the range of $\theta=13^{\rm{o}}\sim 17^{\rm{o}}$ \cite{moon2013optical}. 
There the electrical gating is of particular interest because the gate-driven peak shift/suppression may lead to a possible color-change. 
Further electro-optic application could be found by extending the notion of tunable band structure to other bilayer materials. Recent studies showed that
a 2D material can have many kinds of band structure, linear or parabolic, metallic or insulating, direct or indirect gap, 
and so on \cite{kuc2011influence,cheiwchanchamnangij2012quasiparticle,zahid2013generic,ruppert2014optical,rudenko2014quasiparticle}.
When two such monolayers M1 and M2 form a bilayer composite, M1/M2,  we can assume that the same layer-dependent screening principle applies, i.e., 
M1-band shifts relative to M2-band when gated, which can create new kinds of optical changes other than what we observed in tBG, which will be interesting to investigate theoretically and experimentally.
%In figure~\ref{fig:4} we demonstrate  one such example for A/B = linear-band/insulator-band composite, where a direct-type optical transition is turned on by gating. 
%In this regard various tunable optical transitions may emerge from the numerous A/B library, 

{\it Acknowledgments.}
This work was supported by the National Research Foundation of Korea(NRF) grant funded by the Korea government(MSIT) (No. 2017R1A2B4007782).
P.M. was supported by the NYU Shanghai (start-up funds) and the NYU-ECNU Institute of Physics.
The work at SKKU was supported by the Institute for Basic Science (IBS-R011-D1).

%merlin.mbs apsrev4-1.bst 2010-07-25 4.21a (PWD, AO, DPC) hacked
%Control: key (0)
%Control: author (8) initials jnrlst
%Control: editor formatted (1) identically to author
%Control: production of article title (-1) disabled
%Control: page (0) single
%Control: year (1) truncated
%Control: production of eprint (0) enabled
%

%\bibliography{reference}

\begin{thebibliography}{40}%
\makeatletter
\providecommand \@ifxundefined [1]{%
 \@ifx{#1\undefined}
}%
\providecommand \@ifnum [1]{%
 \ifnum #1\expandafter \@firstoftwo
 \else \expandafter \@secondoftwo
 \fi
}%
\providecommand \@ifx [1]{%
 \ifx #1\expandafter \@firstoftwo
 \else \expandafter \@secondoftwo
 \fi
}%
\providecommand \natexlab [1]{#1}%
\providecommand \enquote  [1]{``#1''}%
\providecommand \bibnamefont  [1]{#1}%
\providecommand \bibfnamefont [1]{#1}%
\providecommand \citenamefont [1]{#1}%
\providecommand \href@noop [0]{\@secondoftwo}%
\providecommand \href [0]{\begingroup \@sanitize@url \@href}%
\providecommand \@href[1]{\@@startlink{#1}\@@href}%
\providecommand \@@href[1]{\endgroup#1\@@endlink}%
\providecommand \@sanitize@url [0]{\catcode `\\12\catcode `\$12\catcode
  `\&12\catcode `\#12\catcode `\^12\catcode `\_12\catcode `\%12\relax}%
\providecommand \@@startlink[1]{}%
\providecommand \@@endlink[0]{}%
\providecommand \url  [0]{\begingroup\@sanitize@url \@url }%
\providecommand \@url [1]{\endgroup\@href {#1}{\urlprefix }}%
\providecommand \urlprefix  [0]{URL }%
\providecommand \Eprint [0]{\href }%
\providecommand \doibase [0]{http://dx.doi.org/}%
\providecommand \selectlanguage [0]{\@gobble}%
\providecommand \bibinfo  [0]{\@secondoftwo}%
\providecommand \bibfield  [0]{\@secondoftwo}%
\providecommand \translation [1]{[#1]}%
\providecommand \BibitemOpen [0]{}%
\providecommand \bibitemStop [0]{}%
\providecommand \bibitemNoStop [0]{.\EOS\space}%
\providecommand \EOS [0]{\spacefactor3000\relax}%
\providecommand \BibitemShut  [1]{\csname bibitem#1\endcsname}%
\let\auto@bib@innerbib\@empty
%</preamble>
\bibitem [{\citenamefont {Novoselov}\ \emph {et~al.}(2004)\citenamefont
  {Novoselov}, \citenamefont {Geim}, \citenamefont {Morozov}, \citenamefont
  {Jiang}, \citenamefont {Zhang}, \citenamefont {Dubonos}, \citenamefont
  {Grigorieva},\ and\ \citenamefont {Firsov}}]{novoselov2004electric}%
  \BibitemOpen
  \bibfield  {author} {\bibinfo {author} {\bibfnamefont {K.~S.}\ \bibnamefont
  {Novoselov}}, \bibinfo {author} {\bibfnamefont {A.~K.}\ \bibnamefont {Geim}},
  \bibinfo {author} {\bibfnamefont {S.~V.}\ \bibnamefont {Morozov}}, \bibinfo
  {author} {\bibfnamefont {D.}~\bibnamefont {Jiang}}, \bibinfo {author}
  {\bibfnamefont {Y.}~\bibnamefont {Zhang}}, \bibinfo {author} {\bibfnamefont
  {S.~V.}\ \bibnamefont {Dubonos}}, \bibinfo {author} {\bibfnamefont {I.~V.}\
  \bibnamefont {Grigorieva}}, \ and\ \bibinfo {author} {\bibfnamefont {A.~A.}\
  \bibnamefont {Firsov}},\ }\href@noop {} {\bibfield  {journal} {\bibinfo
  {journal} {Science}\ }\textbf {\bibinfo {volume} {306}},\ \bibinfo {pages}
  {666} (\bibinfo {year} {2004})}\BibitemShut {NoStop}%
\bibitem [{\citenamefont {Novoselov}\ \emph {et~al.}(2005)\citenamefont
  {Novoselov}, \citenamefont {Geim}, \citenamefont {Morozov}, \citenamefont
  {Jiang}, \citenamefont {Katsnelson}, \citenamefont {Grigorieva},
  \citenamefont {Dubonos},\ and\ \citenamefont {Firsov}}]{novoselov2005two}%
  \BibitemOpen
  \bibfield  {author} {\bibinfo {author} {\bibfnamefont {K.~S.}\ \bibnamefont
  {Novoselov}}, \bibinfo {author} {\bibfnamefont {A.~K.}\ \bibnamefont {Geim}},
  \bibinfo {author} {\bibfnamefont {S.~V.}\ \bibnamefont {Morozov}}, \bibinfo
  {author} {\bibfnamefont {D.}~\bibnamefont {Jiang}}, \bibinfo {author}
  {\bibfnamefont {M.~I.}\ \bibnamefont {Katsnelson}}, \bibinfo {author}
  {\bibfnamefont {I.~V.}\ \bibnamefont {Grigorieva}}, \bibinfo {author}
  {\bibfnamefont {S.~V.}\ \bibnamefont {Dubonos}}, \ and\ \bibinfo {author}
  {\bibfnamefont {A.~A.}\ \bibnamefont {Firsov}},\ }\href@noop {} {\bibfield
  {journal} {\bibinfo  {journal} {Nature}\ }\textbf {\bibinfo {volume} {438}},\
  \bibinfo {pages} {197} (\bibinfo {year} {2005})}\BibitemShut {NoStop}%
\bibitem [{\citenamefont {Oostinga}\ \emph {et~al.}(2008)\citenamefont
  {Oostinga}, \citenamefont {Heersche}, \citenamefont {Liu}, \citenamefont
  {Morpurgo},\ and\ \citenamefont {Vandersypen}}]{oostinga2008gate}%
  \BibitemOpen
  \bibfield  {author} {\bibinfo {author} {\bibfnamefont {J.~B.}\ \bibnamefont
  {Oostinga}}, \bibinfo {author} {\bibfnamefont {H.~B.}\ \bibnamefont
  {Heersche}}, \bibinfo {author} {\bibfnamefont {X.}~\bibnamefont {Liu}},
  \bibinfo {author} {\bibfnamefont {A.~F.}\ \bibnamefont {Morpurgo}}, \ and\
  \bibinfo {author} {\bibfnamefont {L.~M.~K.}\ \bibnamefont {Vandersypen}},\
  }\href@noop {} {\bibfield  {journal} {\bibinfo  {journal} {Nat. Mater.}\
  }\textbf {\bibinfo {volume} {7}},\ \bibinfo {pages} {151} (\bibinfo {year}
  {2008})}\BibitemShut {NoStop}%
\bibitem [{\citenamefont {Furchi}\ \emph {et~al.}(2014)\citenamefont {Furchi},
  \citenamefont {Polyushkin}, \citenamefont {Pospischil},\ and\ \citenamefont
  {Mueller}}]{furchi2014mechanisms}%
  \BibitemOpen
  \bibfield  {author} {\bibinfo {author} {\bibfnamefont {M.~M.}\ \bibnamefont
  {Furchi}}, \bibinfo {author} {\bibfnamefont {D.~K.}\ \bibnamefont
  {Polyushkin}}, \bibinfo {author} {\bibfnamefont {A.}~\bibnamefont
  {Pospischil}}, \ and\ \bibinfo {author} {\bibfnamefont {T.}~\bibnamefont
  {Mueller}},\ }\href@noop {} {\bibfield  {journal} {\bibinfo  {journal} {Nano
  Lett.}\ }\textbf {\bibinfo {volume} {14}},\ \bibinfo {pages} {6165} (\bibinfo
  {year} {2014})}\BibitemShut {NoStop}%
\bibitem [{\citenamefont {Costanzo}\ \emph {et~al.}(2016)\citenamefont
  {Costanzo}, \citenamefont {Jo}, \citenamefont {Berger},\ and\ \citenamefont
  {Morpurgo}}]{costanzo2016gate}%
  \BibitemOpen
  \bibfield  {author} {\bibinfo {author} {\bibfnamefont {D.}~\bibnamefont
  {Costanzo}}, \bibinfo {author} {\bibfnamefont {S.}~\bibnamefont {Jo}},
  \bibinfo {author} {\bibfnamefont {H.}~\bibnamefont {Berger}}, \ and\ \bibinfo
  {author} {\bibfnamefont {A.~F.}\ \bibnamefont {Morpurgo}},\ }\href@noop {}
  {\bibfield  {journal} {\bibinfo  {journal} {Nat. Nanotechnol.}\ }\textbf
  {\bibinfo {volume} {11}},\ \bibinfo {pages} {339} (\bibinfo {year}
  {2016})}\BibitemShut {NoStop}%
\bibitem [{\citenamefont {Nair}\ \emph {et~al.}(2008)\citenamefont {Nair},
  \citenamefont {Blake}, \citenamefont {Grigorenko}, \citenamefont {Novoselov},
  \citenamefont {Booth}, \citenamefont {Stauber}, \citenamefont {Peres},\ and\
  \citenamefont {Geim}}]{nair2008fine}%
  \BibitemOpen
  \bibfield  {author} {\bibinfo {author} {\bibfnamefont {R.~R.}\ \bibnamefont
  {Nair}}, \bibinfo {author} {\bibfnamefont {P.}~\bibnamefont {Blake}},
  \bibinfo {author} {\bibfnamefont {A.~N.}\ \bibnamefont {Grigorenko}},
  \bibinfo {author} {\bibfnamefont {K.~S.}\ \bibnamefont {Novoselov}}, \bibinfo
  {author} {\bibfnamefont {T.~J.}\ \bibnamefont {Booth}}, \bibinfo {author}
  {\bibfnamefont {T.}~\bibnamefont {Stauber}}, \bibinfo {author} {\bibfnamefont
  {N.~M.~R.}\ \bibnamefont {Peres}}, \ and\ \bibinfo {author} {\bibfnamefont
  {A.~K.}\ \bibnamefont {Geim}},\ }\href@noop {} {\bibfield  {journal}
  {\bibinfo  {journal} {Science}\ }\textbf {\bibinfo {volume} {320}},\ \bibinfo
  {pages} {1308} (\bibinfo {year} {2008})}\BibitemShut {NoStop}%
\bibitem [{\citenamefont {Li}\ \emph {et~al.}(2008)\citenamefont {Li},
  \citenamefont {Henriksen}, \citenamefont {Jiang}, \citenamefont {Hao},
  \citenamefont {Martin}, \citenamefont {Kim}, \citenamefont {Stormer},\ and\
  \citenamefont {Basov}}]{li2008dirac}%
  \BibitemOpen
  \bibfield  {author} {\bibinfo {author} {\bibfnamefont {Z.~Q.}\ \bibnamefont
  {Li}}, \bibinfo {author} {\bibfnamefont {E.~A.}\ \bibnamefont {Henriksen}},
  \bibinfo {author} {\bibfnamefont {Z.}~\bibnamefont {Jiang}}, \bibinfo
  {author} {\bibfnamefont {Z.}~\bibnamefont {Hao}}, \bibinfo {author}
  {\bibfnamefont {M.~C.}\ \bibnamefont {Martin}}, \bibinfo {author}
  {\bibfnamefont {P.}~\bibnamefont {Kim}}, \bibinfo {author} {\bibfnamefont
  {H.~L.}\ \bibnamefont {Stormer}}, \ and\ \bibinfo {author} {\bibfnamefont
  {D.~N.}\ \bibnamefont {Basov}},\ }\href@noop {} {\bibfield  {journal}
  {\bibinfo  {journal} {Nat. Phys.}\ }\textbf {\bibinfo {volume} {4}},\
  \bibinfo {pages} {532} (\bibinfo {year} {2008})}\BibitemShut {NoStop}%
\bibitem [{\citenamefont {Mak}\ \emph {et~al.}(2008)\citenamefont {Mak},
  \citenamefont {Sfeir}, \citenamefont {Wu}, \citenamefont {Lui}, \citenamefont
  {Misewich},\ and\ \citenamefont {Heinz}}]{mak2008measurement}%
  \BibitemOpen
  \bibfield  {author} {\bibinfo {author} {\bibfnamefont {K.~F.}\ \bibnamefont
  {Mak}}, \bibinfo {author} {\bibfnamefont {M.~Y.}\ \bibnamefont {Sfeir}},
  \bibinfo {author} {\bibfnamefont {Y.}~\bibnamefont {Wu}}, \bibinfo {author}
  {\bibfnamefont {C.~H.}\ \bibnamefont {Lui}}, \bibinfo {author} {\bibfnamefont
  {J.~A.}\ \bibnamefont {Misewich}}, \ and\ \bibinfo {author} {\bibfnamefont
  {T.~F.}\ \bibnamefont {Heinz}},\ }\href@noop {} {\bibfield  {journal}
  {\bibinfo  {journal} {Phys. Rev. Lett.}\ }\textbf {\bibinfo {volume} {101}},\
  \bibinfo {pages} {196405} (\bibinfo {year} {2008})}\BibitemShut {NoStop}%
\bibitem [{\citenamefont {Jiang}\ \emph {et~al.}(2018)\citenamefont {Jiang},
  \citenamefont {Huang}, \citenamefont {Cheng}, \citenamefont {Fan},
  \citenamefont {Zhang}, \citenamefont {Shan}, \citenamefont {Yi},
  \citenamefont {Dai}, \citenamefont {Shi}, \citenamefont {Liu}, \citenamefont
  {Zeng}, \citenamefont {Zi}, \citenamefont {Sipe}, \citenamefont {Shen},
  \citenamefont {Liu},\ and\ \citenamefont {Wu}}]{jiang2018gate}%
  \BibitemOpen
  \bibfield  {author} {\bibinfo {author} {\bibfnamefont {T.}~\bibnamefont
  {Jiang}}, \bibinfo {author} {\bibfnamefont {D.}~\bibnamefont {Huang}},
  \bibinfo {author} {\bibfnamefont {J.}~\bibnamefont {Cheng}}, \bibinfo
  {author} {\bibfnamefont {X.}~\bibnamefont {Fan}}, \bibinfo {author}
  {\bibfnamefont {Z.}~\bibnamefont {Zhang}}, \bibinfo {author} {\bibfnamefont
  {Y.}~\bibnamefont {Shan}}, \bibinfo {author} {\bibfnamefont {Y.}~\bibnamefont
  {Yi}}, \bibinfo {author} {\bibfnamefont {Y.}~\bibnamefont {Dai}}, \bibinfo
  {author} {\bibfnamefont {L.}~\bibnamefont {Shi}}, \bibinfo {author}
  {\bibfnamefont {K.}~\bibnamefont {Liu}}, \bibinfo {author} {\bibfnamefont
  {C.}~\bibnamefont {Zeng}}, \bibinfo {author} {\bibfnamefont {J.}~\bibnamefont
  {Zi}}, \bibinfo {author} {\bibfnamefont {J.~E.}\ \bibnamefont {Sipe}},
  \bibinfo {author} {\bibfnamefont {Y.-R.}\ \bibnamefont {Shen}}, \bibinfo
  {author} {\bibfnamefont {W.-T.}\ \bibnamefont {Liu}}, \ and\ \bibinfo
  {author} {\bibfnamefont {S.~o.}\ \bibnamefont {Wu}},\ }\href@noop {}
  {\bibfield  {journal} {\bibinfo  {journal} {Nat. Photonics}\ }\textbf
  {\bibinfo {volume} {12}},\ \bibinfo {pages} {430} (\bibinfo {year}
  {2018})}\BibitemShut {NoStop}%
\bibitem [{\citenamefont {Liu}\ \emph {et~al.}(2011)\citenamefont {Liu},
  \citenamefont {Yin}, \citenamefont {Ulin-Avila}, \citenamefont {Geng},
  \citenamefont {Zentgraf}, \citenamefont {Ju}, \citenamefont {Wang},\ and\
  \citenamefont {Zhang}}]{liu2011graphene}%
  \BibitemOpen
  \bibfield  {author} {\bibinfo {author} {\bibfnamefont {M.}~\bibnamefont
  {Liu}}, \bibinfo {author} {\bibfnamefont {X.}~\bibnamefont {Yin}}, \bibinfo
  {author} {\bibfnamefont {E.}~\bibnamefont {Ulin-Avila}}, \bibinfo {author}
  {\bibfnamefont {B.}~\bibnamefont {Geng}}, \bibinfo {author} {\bibfnamefont
  {T.}~\bibnamefont {Zentgraf}}, \bibinfo {author} {\bibfnamefont
  {L.}~\bibnamefont {Ju}}, \bibinfo {author} {\bibfnamefont {F.}~\bibnamefont
  {Wang}}, \ and\ \bibinfo {author} {\bibfnamefont {X.}~\bibnamefont {Zhang}},\
  }\href@noop {} {\bibfield  {journal} {\bibinfo  {journal} {Nature}\ }\textbf
  {\bibinfo {volume} {474}},\ \bibinfo {pages} {64} (\bibinfo {year}
  {2011})}\BibitemShut {NoStop}%
\bibitem [{\citenamefont {Zhang}\ \emph {et~al.}(2009)\citenamefont {Zhang},
  \citenamefont {Tang}, \citenamefont {Girit}, \citenamefont {Hao},
  \citenamefont {Martin}, \citenamefont {Zettl}, \citenamefont {Crommie},
  \citenamefont {Shen},\ and\ \citenamefont {Wang}}]{zhang2009direct}%
  \BibitemOpen
  \bibfield  {author} {\bibinfo {author} {\bibfnamefont {Y.}~\bibnamefont
  {Zhang}}, \bibinfo {author} {\bibfnamefont {T.-T.}\ \bibnamefont {Tang}},
  \bibinfo {author} {\bibfnamefont {C.}~\bibnamefont {Girit}}, \bibinfo
  {author} {\bibfnamefont {Z.}~\bibnamefont {Hao}}, \bibinfo {author}
  {\bibfnamefont {M.~C.}\ \bibnamefont {Martin}}, \bibinfo {author}
  {\bibfnamefont {A.}~\bibnamefont {Zettl}}, \bibinfo {author} {\bibfnamefont
  {M.~F.}\ \bibnamefont {Crommie}}, \bibinfo {author} {\bibfnamefont {Y.~R.}\
  \bibnamefont {Shen}}, \ and\ \bibinfo {author} {\bibfnamefont
  {F.}~\bibnamefont {Wang}},\ }\href@noop {} {\bibfield  {journal} {\bibinfo
  {journal} {Nature}\ }\textbf {\bibinfo {volume} {459}},\ \bibinfo {pages}
  {820} (\bibinfo {year} {2009})}\BibitemShut {NoStop}%
\bibitem [{\citenamefont {Li}\ \emph {et~al.}(2009)\citenamefont {Li},
  \citenamefont {Henriksen}, \citenamefont {Jiang}, \citenamefont {Hao},
  \citenamefont {Martin}, \citenamefont {Kim}, \citenamefont {Stormer},\ and\
  \citenamefont {Basov}}]{li2009band}%
  \BibitemOpen
  \bibfield  {author} {\bibinfo {author} {\bibfnamefont {Z.~Q.}\ \bibnamefont
  {Li}}, \bibinfo {author} {\bibfnamefont {E.~A.}\ \bibnamefont {Henriksen}},
  \bibinfo {author} {\bibfnamefont {Z.}~\bibnamefont {Jiang}}, \bibinfo
  {author} {\bibfnamefont {Z.}~\bibnamefont {Hao}}, \bibinfo {author}
  {\bibfnamefont {M.~C.}\ \bibnamefont {Martin}}, \bibinfo {author}
  {\bibfnamefont {P.}~\bibnamefont {Kim}}, \bibinfo {author} {\bibfnamefont
  {H.~L.}\ \bibnamefont {Stormer}}, \ and\ \bibinfo {author} {\bibfnamefont
  {D.~N.}\ \bibnamefont {Basov}},\ }\href@noop {} {\bibfield  {journal}
  {\bibinfo  {journal} {Phys. Rev. Lett.}\ }\textbf {\bibinfo {volume} {102}},\
  \bibinfo {pages} {037403} (\bibinfo {year} {2009})}\BibitemShut {NoStop}%
\bibitem [{\citenamefont {Chernikov}\ \emph {et~al.}(2015)\citenamefont
  {Chernikov}, \citenamefont {van~der Zande}, \citenamefont {Hill},
  \citenamefont {Rigosi}, \citenamefont {Velauthapillai}, \citenamefont
  {Hone},\ and\ \citenamefont {Heinz}}]{chernikov2015electrical}%
  \BibitemOpen
  \bibfield  {author} {\bibinfo {author} {\bibfnamefont {A.}~\bibnamefont
  {Chernikov}}, \bibinfo {author} {\bibfnamefont {A.~M.}\ \bibnamefont {van~der
  Zande}}, \bibinfo {author} {\bibfnamefont {H.~M.}\ \bibnamefont {Hill}},
  \bibinfo {author} {\bibfnamefont {A.~F.}\ \bibnamefont {Rigosi}}, \bibinfo
  {author} {\bibfnamefont {A.}~\bibnamefont {Velauthapillai}}, \bibinfo
  {author} {\bibfnamefont {J.}~\bibnamefont {Hone}}, \ and\ \bibinfo {author}
  {\bibfnamefont {T.~F.}\ \bibnamefont {Heinz}},\ }\href@noop {} {\bibfield
  {journal} {\bibinfo  {journal} {Phys. Rev. Lett.}\ }\textbf {\bibinfo
  {volume} {115}},\ \bibinfo {pages} {126802} (\bibinfo {year}
  {2015})}\BibitemShut {NoStop}%
\bibitem [{\citenamefont {Whitney}\ \emph {et~al.}(2016)\citenamefont
  {Whitney}, \citenamefont {Sherrott}, \citenamefont {Jariwala}, \citenamefont
  {Lin}, \citenamefont {Bechtel}, \citenamefont {Rossman},\ and\ \citenamefont
  {Atwater}}]{whitney2016field}%
  \BibitemOpen
  \bibfield  {author} {\bibinfo {author} {\bibfnamefont {W.~S.}\ \bibnamefont
  {Whitney}}, \bibinfo {author} {\bibfnamefont {M.~C.}\ \bibnamefont
  {Sherrott}}, \bibinfo {author} {\bibfnamefont {D.}~\bibnamefont {Jariwala}},
  \bibinfo {author} {\bibfnamefont {W.-H.}\ \bibnamefont {Lin}}, \bibinfo
  {author} {\bibfnamefont {H.~A.}\ \bibnamefont {Bechtel}}, \bibinfo {author}
  {\bibfnamefont {G.~R.}\ \bibnamefont {Rossman}}, \ and\ \bibinfo {author}
  {\bibfnamefont {H.~A.}\ \bibnamefont {Atwater}},\ }\href@noop {} {\bibfield
  {journal} {\bibinfo  {journal} {Nano Lett.}\ }\textbf {\bibinfo {volume}
  {17}},\ \bibinfo {pages} {78} (\bibinfo {year} {2016})}\BibitemShut {NoStop}%
\bibitem [{\citenamefont {Ju}\ \emph {et~al.}(2017)\citenamefont {Ju},
  \citenamefont {Wang}, \citenamefont {Cao}, \citenamefont {Taniguchi},
  \citenamefont {Watanabe}, \citenamefont {Louie}, \citenamefont {Rana},
  \citenamefont {Park}, \citenamefont {Hone}, \citenamefont {Wang},\ and\
  \citenamefont {McEuen}}]{ju2017tunable}%
  \BibitemOpen
  \bibfield  {author} {\bibinfo {author} {\bibfnamefont {L.}~\bibnamefont
  {Ju}}, \bibinfo {author} {\bibfnamefont {L.}~\bibnamefont {Wang}}, \bibinfo
  {author} {\bibfnamefont {T.}~\bibnamefont {Cao}}, \bibinfo {author}
  {\bibfnamefont {T.}~\bibnamefont {Taniguchi}}, \bibinfo {author}
  {\bibfnamefont {K.}~\bibnamefont {Watanabe}}, \bibinfo {author}
  {\bibfnamefont {S.~G.}\ \bibnamefont {Louie}}, \bibinfo {author}
  {\bibfnamefont {F.}~\bibnamefont {Rana}}, \bibinfo {author} {\bibfnamefont
  {J.}~\bibnamefont {Park}}, \bibinfo {author} {\bibfnamefont {J.}~\bibnamefont
  {Hone}}, \bibinfo {author} {\bibfnamefont {F.}~\bibnamefont {Wang}}, \ and\
  \bibinfo {author} {\bibfnamefont {P.~L.}\ \bibnamefont {McEuen}},\
  }\href@noop {} {\bibfield  {journal} {\bibinfo  {journal} {Science}\ }\textbf
  {\bibinfo {volume} {358}},\ \bibinfo {pages} {907} (\bibinfo {year}
  {2017})}\BibitemShut {NoStop}%
\bibitem [{\citenamefont {Yao}\ \emph {et~al.}(2017)\citenamefont {Yao},
  \citenamefont {Yan}, \citenamefont {Kahn}, \citenamefont {Suslu},
  \citenamefont {Liang}, \citenamefont {Barnard}, \citenamefont {Tongay},
  \citenamefont {Zettl}, \citenamefont {Borys},\ and\ \citenamefont
  {Schuck}}]{yao2017optically}%
  \BibitemOpen
  \bibfield  {author} {\bibinfo {author} {\bibfnamefont {K.}~\bibnamefont
  {Yao}}, \bibinfo {author} {\bibfnamefont {A.}~\bibnamefont {Yan}}, \bibinfo
  {author} {\bibfnamefont {S.}~\bibnamefont {Kahn}}, \bibinfo {author}
  {\bibfnamefont {A.}~\bibnamefont {Suslu}}, \bibinfo {author} {\bibfnamefont
  {Y.}~\bibnamefont {Liang}}, \bibinfo {author} {\bibfnamefont {E.~S.}\
  \bibnamefont {Barnard}}, \bibinfo {author} {\bibfnamefont {S.}~\bibnamefont
  {Tongay}}, \bibinfo {author} {\bibfnamefont {A.}~\bibnamefont {Zettl}},
  \bibinfo {author} {\bibfnamefont {N.~J.}\ \bibnamefont {Borys}}, \ and\
  \bibinfo {author} {\bibfnamefont {P.~J.}\ \bibnamefont {Schuck}},\
  }\href@noop {} {\bibfield  {journal} {\bibinfo  {journal} {Phys. Rev. Lett.}\
  }\textbf {\bibinfo {volume} {119}},\ \bibinfo {pages} {087401} (\bibinfo
  {year} {2017})}\BibitemShut {NoStop}%
\bibitem [{\citenamefont {Ohta}\ \emph {et~al.}(2012)\citenamefont {Ohta},
  \citenamefont {Robinson}, \citenamefont {Feibelman}, \citenamefont
  {Bostwick}, \citenamefont {Rotenberg},\ and\ \citenamefont
  {Beechem}}]{ohta2012evidence}%
  \BibitemOpen
  \bibfield  {author} {\bibinfo {author} {\bibfnamefont {T.}~\bibnamefont
  {Ohta}}, \bibinfo {author} {\bibfnamefont {J.~T.}\ \bibnamefont {Robinson}},
  \bibinfo {author} {\bibfnamefont {P.~J.}\ \bibnamefont {Feibelman}}, \bibinfo
  {author} {\bibfnamefont {A.}~\bibnamefont {Bostwick}}, \bibinfo {author}
  {\bibfnamefont {E.}~\bibnamefont {Rotenberg}}, \ and\ \bibinfo {author}
  {\bibfnamefont {T.~E.}\ \bibnamefont {Beechem}},\ }\href@noop {} {\bibfield
  {journal} {\bibinfo  {journal} {Phys. Rev. Lett.}\ }\textbf {\bibinfo
  {volume} {109}},\ \bibinfo {pages} {186807} (\bibinfo {year}
  {2012})}\BibitemShut {NoStop}%
\bibitem [{\citenamefont {Bistritzer}\ and\ \citenamefont
  {MacDonald}(2011)}]{bistritzer2011moire}%
  \BibitemOpen
  \bibfield  {author} {\bibinfo {author} {\bibfnamefont {R.}~\bibnamefont
  {Bistritzer}}\ and\ \bibinfo {author} {\bibfnamefont {A.~H.}\ \bibnamefont
  {MacDonald}},\ }\href@noop {} {\bibfield  {journal} {\bibinfo  {journal}
  {Phys. Rev. B}\ }\textbf {\bibinfo {volume} {84}},\ \bibinfo {pages} {035440}
  (\bibinfo {year} {2011})}\BibitemShut {NoStop}%
\bibitem [{\citenamefont {Moon}\ and\ \citenamefont
  {Koshino}(2012)}]{moon2012energy}%
  \BibitemOpen
  \bibfield  {author} {\bibinfo {author} {\bibfnamefont {P.}~\bibnamefont
  {Moon}}\ and\ \bibinfo {author} {\bibfnamefont {M.}~\bibnamefont {Koshino}},\
  }\href@noop {} {\bibfield  {journal} {\bibinfo  {journal} {Phys. Rev. B}\
  }\textbf {\bibinfo {volume} {85}},\ \bibinfo {pages} {195458} (\bibinfo
  {year} {2012})}\BibitemShut {NoStop}%
\bibitem [{\citenamefont {Cao}\ \emph {et~al.}(2018)\citenamefont {Cao},
  \citenamefont {Fatemi}, \citenamefont {Fang}, \citenamefont {Watanabe},
  \citenamefont {Taniguchi}, \citenamefont {Kaxiras},\ and\ \citenamefont
  {Jarillo-Herrero}}]{cao2018unconventional}%
  \BibitemOpen
  \bibfield  {author} {\bibinfo {author} {\bibfnamefont {Y.}~\bibnamefont
  {Cao}}, \bibinfo {author} {\bibfnamefont {V.}~\bibnamefont {Fatemi}},
  \bibinfo {author} {\bibfnamefont {S.}~\bibnamefont {Fang}}, \bibinfo {author}
  {\bibfnamefont {K.}~\bibnamefont {Watanabe}}, \bibinfo {author}
  {\bibfnamefont {T.}~\bibnamefont {Taniguchi}}, \bibinfo {author}
  {\bibfnamefont {E.}~\bibnamefont {Kaxiras}}, \ and\ \bibinfo {author}
  {\bibfnamefont {P.}~\bibnamefont {Jarillo-Herrero}},\ }\href@noop {}
  {\bibfield  {journal} {\bibinfo  {journal} {Nature}\ }\textbf {\bibinfo
  {volume} {556}},\ \bibinfo {pages} {43} (\bibinfo {year} {2018})}\BibitemShut
  {NoStop}%
\bibitem [{\citenamefont {dos Santos}\ \emph {et~al.}(2007)\citenamefont {dos
  Santos}, \citenamefont {Peres},\ and\ \citenamefont
  {Neto}}]{dos2007graphene}%
  \BibitemOpen
  \bibfield  {author} {\bibinfo {author} {\bibfnamefont {J.~M. B.~L.}\
  \bibnamefont {dos Santos}}, \bibinfo {author} {\bibfnamefont {N.~M.~R.}\
  \bibnamefont {Peres}}, \ and\ \bibinfo {author} {\bibfnamefont {A.~H.~C.}\
  \bibnamefont {Neto}},\ }\href@noop {} {\bibfield  {journal} {\bibinfo
  {journal} {Phys. Rev. Lett.}\ }\textbf {\bibinfo {volume} {99}},\ \bibinfo
  {pages} {256802} (\bibinfo {year} {2007})}\BibitemShut {NoStop}%
\bibitem [{\citenamefont {Mele}(2010)}]{mele2010commensuration}%
  \BibitemOpen
  \bibfield  {author} {\bibinfo {author} {\bibfnamefont {E.~J.}\ \bibnamefont
  {Mele}},\ }\href@noop {} {\bibfield  {journal} {\bibinfo  {journal} {Phys.
  Rev. B}\ }\textbf {\bibinfo {volume} {81}},\ \bibinfo {pages} {161405(R)}
  (\bibinfo {year} {2010})}\BibitemShut {NoStop}%
\bibitem [{\citenamefont {dos Santos}\ \emph {et~al.}(2012)\citenamefont {dos
  Santos}, \citenamefont {Peres},\ and\ \citenamefont
  {Neto}}]{dos2012continuum}%
  \BibitemOpen
  \bibfield  {author} {\bibinfo {author} {\bibfnamefont {J.~M. B.~L.}\
  \bibnamefont {dos Santos}}, \bibinfo {author} {\bibfnamefont {N.~M.~R.}\
  \bibnamefont {Peres}}, \ and\ \bibinfo {author} {\bibfnamefont {A.~H.~C.}\
  \bibnamefont {Neto}},\ }\href@noop {} {\bibfield  {journal} {\bibinfo
  {journal} {Phys. Rev. B}\ }\textbf {\bibinfo {volume} {86}},\ \bibinfo
  {pages} {155449} (\bibinfo {year} {2012})}\BibitemShut {NoStop}%
\bibitem [{\citenamefont {Tabert}\ and\ \citenamefont
  {Nicol}(2013)}]{tabert2013optical}%
  \BibitemOpen
  \bibfield  {author} {\bibinfo {author} {\bibfnamefont {C.~J.}\ \bibnamefont
  {Tabert}}\ and\ \bibinfo {author} {\bibfnamefont {E.~J.}\ \bibnamefont
  {Nicol}},\ }\href@noop {} {\bibfield  {journal} {\bibinfo  {journal} {Phys.
  Rev. B}\ }\textbf {\bibinfo {volume} {87}},\ \bibinfo {pages} {121402(R)}
  (\bibinfo {year} {2013})}\BibitemShut {NoStop}%
\bibitem [{\citenamefont {Moon}\ and\ \citenamefont
  {Koshino}(2013)}]{moon2013optical}%
  \BibitemOpen
  \bibfield  {author} {\bibinfo {author} {\bibfnamefont {P.}~\bibnamefont
  {Moon}}\ and\ \bibinfo {author} {\bibfnamefont {M.}~\bibnamefont {Koshino}},\
  }\href@noop {} {\bibfield  {journal} {\bibinfo  {journal} {Phys. Rev. B}\
  }\textbf {\bibinfo {volume} {87}},\ \bibinfo {pages} {205404} (\bibinfo
  {year} {2013})}\BibitemShut {NoStop}%
\bibitem [{\citenamefont {Havener}\ \emph {et~al.}(2014)\citenamefont
  {Havener}, \citenamefont {Liang}, \citenamefont {Brown}, \citenamefont
  {Yang},\ and\ \citenamefont {Park}}]{havener2014van}%
  \BibitemOpen
  \bibfield  {author} {\bibinfo {author} {\bibfnamefont {R.~W.}\ \bibnamefont
  {Havener}}, \bibinfo {author} {\bibfnamefont {Y.}~\bibnamefont {Liang}},
  \bibinfo {author} {\bibfnamefont {L.}~\bibnamefont {Brown}}, \bibinfo
  {author} {\bibfnamefont {L.}~\bibnamefont {Yang}}, \ and\ \bibinfo {author}
  {\bibfnamefont {J.}~\bibnamefont {Park}},\ }\href@noop {} {\bibfield
  {journal} {\bibinfo  {journal} {Nano Lett.}\ }\textbf {\bibinfo {volume}
  {14}},\ \bibinfo {pages} {3353} (\bibinfo {year} {2014})}\BibitemShut
  {NoStop}%
\bibitem [{\citenamefont {Kim}\ \emph {et~al.}(2016)\citenamefont {Kim},
  \citenamefont {S{\'a}nchez-Castillo}, \citenamefont {Ziegler}, \citenamefont
  {Ogawa}, \citenamefont {Noguez},\ and\ \citenamefont {Park}}]{kim2016chiral}%
  \BibitemOpen
  \bibfield  {author} {\bibinfo {author} {\bibfnamefont {C.-J.}\ \bibnamefont
  {Kim}}, \bibinfo {author} {\bibfnamefont {A.}~\bibnamefont
  {S{\'a}nchez-Castillo}}, \bibinfo {author} {\bibfnamefont {Z.}~\bibnamefont
  {Ziegler}}, \bibinfo {author} {\bibfnamefont {Y.}~\bibnamefont {Ogawa}},
  \bibinfo {author} {\bibfnamefont {C.}~\bibnamefont {Noguez}}, \ and\ \bibinfo
  {author} {\bibfnamefont {J.}~\bibnamefont {Park}},\ }\href@noop {} {\bibfield
   {journal} {\bibinfo  {journal} {Nat. Nanotechnol.}\ }\textbf {\bibinfo
  {volume} {11}},\ \bibinfo {pages} {520} (\bibinfo {year} {2016})}\BibitemShut
  {NoStop}%
\bibitem [{\citenamefont {Nguyen}\ \emph {et~al.}(2015)\citenamefont {Nguyen},
  \citenamefont {Shin}, \citenamefont {Duong}, \citenamefont {Kim},
  \citenamefont {Perello}, \citenamefont {Lim}, \citenamefont {Yuan},
  \citenamefont {Ding}, \citenamefont {Jeong}, \citenamefont {Shin},
  \citenamefont {Lee}, \citenamefont {Chae}, \citenamefont {Vu}, \citenamefont
  {Lee},\ and\ \citenamefont {Lee}}]{nguyen2015seamless}%
  \BibitemOpen
  \bibfield  {author} {\bibinfo {author} {\bibfnamefont {V.~L.}\ \bibnamefont
  {Nguyen}}, \bibinfo {author} {\bibfnamefont {B.~G.}\ \bibnamefont {Shin}},
  \bibinfo {author} {\bibfnamefont {D.~L.}\ \bibnamefont {Duong}}, \bibinfo
  {author} {\bibfnamefont {S.~T.}\ \bibnamefont {Kim}}, \bibinfo {author}
  {\bibfnamefont {D.}~\bibnamefont {Perello}}, \bibinfo {author} {\bibfnamefont
  {Y.~J.}\ \bibnamefont {Lim}}, \bibinfo {author} {\bibfnamefont {Q.~H.}\
  \bibnamefont {Yuan}}, \bibinfo {author} {\bibfnamefont {F.}~\bibnamefont
  {Ding}}, \bibinfo {author} {\bibfnamefont {H.~Y.}\ \bibnamefont {Jeong}},
  \bibinfo {author} {\bibfnamefont {H.~S.}\ \bibnamefont {Shin}}, \bibinfo
  {author} {\bibfnamefont {S.~M.}\ \bibnamefont {Lee}}, \bibinfo {author}
  {\bibfnamefont {S.~H.}\ \bibnamefont {Chae}}, \bibinfo {author}
  {\bibfnamefont {Q.~A.}\ \bibnamefont {Vu}}, \bibinfo {author} {\bibfnamefont
  {S.~H.}\ \bibnamefont {Lee}}, \ and\ \bibinfo {author} {\bibfnamefont
  {Y.~H.}\ \bibnamefont {Lee}},\ }\href@noop {} {\bibfield  {journal} {\bibinfo
   {journal} {Adv. Mater.}\ }\textbf {\bibinfo {volume} {27}},\ \bibinfo
  {pages} {1376} (\bibinfo {year} {2015})}\BibitemShut {NoStop}%
\bibitem [{\citenamefont {Nguyen}\ \emph {et~al.}(2016)\citenamefont {Nguyen},
  \citenamefont {Perello}, \citenamefont {Lee}, \citenamefont {Nai},
  \citenamefont {Shin}, \citenamefont {Kim}, \citenamefont {Park},
  \citenamefont {Jeong}, \citenamefont {Zhao}, \citenamefont {Vu},
  \citenamefont {Lee}, \citenamefont {Loh}, \citenamefont {Jeong},\ and\
  \citenamefont {Lee}}]{nguyen2016wafer}%
  \BibitemOpen
  \bibfield  {author} {\bibinfo {author} {\bibfnamefont {V.~L.}\ \bibnamefont
  {Nguyen}}, \bibinfo {author} {\bibfnamefont {D.~J.}\ \bibnamefont {Perello}},
  \bibinfo {author} {\bibfnamefont {S.}~\bibnamefont {Lee}}, \bibinfo {author}
  {\bibfnamefont {C.~T.}\ \bibnamefont {Nai}}, \bibinfo {author} {\bibfnamefont
  {B.~G.}\ \bibnamefont {Shin}}, \bibinfo {author} {\bibfnamefont {J.-G.}\
  \bibnamefont {Kim}}, \bibinfo {author} {\bibfnamefont {H.~Y.}\ \bibnamefont
  {Park}}, \bibinfo {author} {\bibfnamefont {H.~Y.}\ \bibnamefont {Jeong}},
  \bibinfo {author} {\bibfnamefont {J.}~\bibnamefont {Zhao}}, \bibinfo {author}
  {\bibfnamefont {Q.~A.}\ \bibnamefont {Vu}}, \bibinfo {author} {\bibfnamefont
  {S.~H.}\ \bibnamefont {Lee}}, \bibinfo {author} {\bibfnamefont {K.~P.}\
  \bibnamefont {Loh}}, \bibinfo {author} {\bibfnamefont {S.-Y.}\ \bibnamefont
  {Jeong}}, \ and\ \bibinfo {author} {\bibfnamefont {Y.~H.}\ \bibnamefont
  {Lee}},\ }\href@noop {} {\bibfield  {journal} {\bibinfo  {journal} {Adv.
  Mater.}\ }\textbf {\bibinfo {volume} {28}},\ \bibinfo {pages} {8177}
  (\bibinfo {year} {2016})}\BibitemShut {NoStop}%
\bibitem [{\citenamefont {Horng}\ \emph {et~al.}(2011)\citenamefont {Horng},
  \citenamefont {Chen}, \citenamefont {Geng}, \citenamefont {Girit},
  \citenamefont {Zhang}, \citenamefont {Hao}, \citenamefont {Bechtel},
  \citenamefont {Martin}, \citenamefont {Zettl}, \citenamefont {Crommie},
  \citenamefont {Shen},\ and\ \citenamefont {Wang}}]{horng2011drude}%
  \BibitemOpen
  \bibfield  {author} {\bibinfo {author} {\bibfnamefont {J.}~\bibnamefont
  {Horng}}, \bibinfo {author} {\bibfnamefont {C.-F.}\ \bibnamefont {Chen}},
  \bibinfo {author} {\bibfnamefont {B.}~\bibnamefont {Geng}}, \bibinfo {author}
  {\bibfnamefont {C.}~\bibnamefont {Girit}}, \bibinfo {author} {\bibfnamefont
  {Y.}~\bibnamefont {Zhang}}, \bibinfo {author} {\bibfnamefont
  {Z.}~\bibnamefont {Hao}}, \bibinfo {author} {\bibfnamefont {H.~A.}\
  \bibnamefont {Bechtel}}, \bibinfo {author} {\bibfnamefont {M.}~\bibnamefont
  {Martin}}, \bibinfo {author} {\bibfnamefont {A.}~\bibnamefont {Zettl}},
  \bibinfo {author} {\bibfnamefont {M.~F.}\ \bibnamefont {Crommie}}, \bibinfo
  {author} {\bibfnamefont {Y.~R.}\ \bibnamefont {Shen}}, \ and\ \bibinfo
  {author} {\bibfnamefont {F.}~\bibnamefont {Wang}},\ }\href@noop {} {\bibfield
   {journal} {\bibinfo  {journal} {Phys. Rev. B}\ }\textbf {\bibinfo {volume}
  {83}},\ \bibinfo {pages} {165113} (\bibinfo {year} {2011})}\BibitemShut
  {NoStop}%
\bibitem [{\citenamefont {Kim}\ \emph {et~al.}(2010)\citenamefont {Kim},
  \citenamefont {Jang}, \citenamefont {Lee}, \citenamefont {Hong},
  \citenamefont {Ahn},\ and\ \citenamefont {Cho}}]{kim2010high}%
  \BibitemOpen
  \bibfield  {author} {\bibinfo {author} {\bibfnamefont {B.~J.}\ \bibnamefont
  {Kim}}, \bibinfo {author} {\bibfnamefont {H.}~\bibnamefont {Jang}}, \bibinfo
  {author} {\bibfnamefont {S.-K.}\ \bibnamefont {Lee}}, \bibinfo {author}
  {\bibfnamefont {B.~H.}\ \bibnamefont {Hong}}, \bibinfo {author}
  {\bibfnamefont {J.-H.}\ \bibnamefont {Ahn}}, \ and\ \bibinfo {author}
  {\bibfnamefont {J.~H.}\ \bibnamefont {Cho}},\ }\href@noop {} {\bibfield
  {journal} {\bibinfo  {journal} {Nano Lett.}\ }\textbf {\bibinfo {volume}
  {10}},\ \bibinfo {pages} {3464} (\bibinfo {year} {2010})}\BibitemShut
  {NoStop}%
\bibitem [{\citenamefont {Kuzmenko}(2005)}]{kuzmenko2005kramers}%
  \BibitemOpen
  \bibfield  {author} {\bibinfo {author} {\bibfnamefont {A.~B.}\ \bibnamefont
  {Kuzmenko}},\ }\href@noop {} {\bibfield  {journal} {\bibinfo  {journal} {Rev.
  Sci. Instrum.}\ }\textbf {\bibinfo {volume} {76}},\ \bibinfo {pages} {083108}
  (\bibinfo {year} {2005})}\BibitemShut {NoStop}%
\bibitem [{\citenamefont {Lee}\ \emph {et~al.}(2011)\citenamefont {Lee},
  \citenamefont {Kim}, \citenamefont {Bae}, \citenamefont {Kim}, \citenamefont
  {Hong},\ and\ \citenamefont {Choi}}]{lee2011optical}%
  \BibitemOpen
  \bibfield  {author} {\bibinfo {author} {\bibfnamefont {C.}~\bibnamefont
  {Lee}}, \bibinfo {author} {\bibfnamefont {J.~Y.}\ \bibnamefont {Kim}},
  \bibinfo {author} {\bibfnamefont {S.}~\bibnamefont {Bae}}, \bibinfo {author}
  {\bibfnamefont {K.~S.}\ \bibnamefont {Kim}}, \bibinfo {author} {\bibfnamefont
  {B.~H.}\ \bibnamefont {Hong}}, \ and\ \bibinfo {author} {\bibfnamefont
  {E.~J.}\ \bibnamefont {Choi}},\ }\href@noop {} {\bibfield  {journal}
  {\bibinfo  {journal} {Appl. Phys. Lett.}\ }\textbf {\bibinfo {volume} {98}},\
  \bibinfo {pages} {071905} (\bibinfo {year} {2011})}\BibitemShut {NoStop}%
\bibitem [{\citenamefont {Yu}\ \emph {et~al.}(2016)\citenamefont {Yu},
  \citenamefont {Kim}, \citenamefont {Kim}, \citenamefont {Lee}, \citenamefont
  {Hwang}, \citenamefont {Hwang},\ and\ \citenamefont {Choi}}]{yu2016infrared}%
  \BibitemOpen
  \bibfield  {author} {\bibinfo {author} {\bibfnamefont {K.}~\bibnamefont
  {Yu}}, \bibinfo {author} {\bibfnamefont {J.}~\bibnamefont {Kim}}, \bibinfo
  {author} {\bibfnamefont {J.~Y.}\ \bibnamefont {Kim}}, \bibinfo {author}
  {\bibfnamefont {W.}~\bibnamefont {Lee}}, \bibinfo {author} {\bibfnamefont
  {J.~Y.}\ \bibnamefont {Hwang}}, \bibinfo {author} {\bibfnamefont {E.~H.}\
  \bibnamefont {Hwang}}, \ and\ \bibinfo {author} {\bibfnamefont {E.~J.}\
  \bibnamefont {Choi}},\ }\href@noop {} {\bibfield  {journal} {\bibinfo
  {journal} {Phys. Rev. B}\ }\textbf {\bibinfo {volume} {94}},\ \bibinfo
  {pages} {235404} (\bibinfo {year} {2016})}\BibitemShut {NoStop}%
\bibitem [{\citenamefont {Moon}\ \emph {et~al.}(2014)\citenamefont {Moon},
  \citenamefont {Son},\ and\ \citenamefont {Koshino}}]{moon2014optical}%
  \BibitemOpen
  \bibfield  {author} {\bibinfo {author} {\bibfnamefont {P.}~\bibnamefont
  {Moon}}, \bibinfo {author} {\bibfnamefont {Y.-W.}\ \bibnamefont {Son}}, \
  and\ \bibinfo {author} {\bibfnamefont {M.}~\bibnamefont {Koshino}},\
  }\href@noop {} {\bibfield  {journal} {\bibinfo  {journal} {Phys. Rev. B}\
  }\textbf {\bibinfo {volume} {90}},\ \bibinfo {pages} {155427} (\bibinfo
  {year} {2014})}\BibitemShut {NoStop}%
\bibitem [{\citenamefont {Kuc}\ \emph {et~al.}(2011)\citenamefont {Kuc},
  \citenamefont {Zibouche},\ and\ \citenamefont {Heine}}]{kuc2011influence}%
  \BibitemOpen
  \bibfield  {author} {\bibinfo {author} {\bibfnamefont {A.}~\bibnamefont
  {Kuc}}, \bibinfo {author} {\bibfnamefont {N.}~\bibnamefont {Zibouche}}, \
  and\ \bibinfo {author} {\bibfnamefont {T.}~\bibnamefont {Heine}},\
  }\href@noop {} {\bibfield  {journal} {\bibinfo  {journal} {Phys. Rev. B}\
  }\textbf {\bibinfo {volume} {83}},\ \bibinfo {pages} {245213} (\bibinfo
  {year} {2011})}\BibitemShut {NoStop}%
\bibitem [{\citenamefont {Cheiwchanchamnangij}\ and\ \citenamefont
  {Lambrecht}(2012)}]{cheiwchanchamnangij2012quasiparticle}%
  \BibitemOpen
  \bibfield  {author} {\bibinfo {author} {\bibfnamefont {T.}~\bibnamefont
  {Cheiwchanchamnangij}}\ and\ \bibinfo {author} {\bibfnamefont {W.~R.~L.}\
  \bibnamefont {Lambrecht}},\ }\href@noop {} {\bibfield  {journal} {\bibinfo
  {journal} {Phys. Rev. B}\ }\textbf {\bibinfo {volume} {85}},\ \bibinfo
  {pages} {205302} (\bibinfo {year} {2012})}\BibitemShut {NoStop}%
\bibitem [{\citenamefont {Zahid}\ \emph {et~al.}(2013)\citenamefont {Zahid},
  \citenamefont {Liu}, \citenamefont {Zhu}, \citenamefont {Wang},\ and\
  \citenamefont {Guo}}]{zahid2013generic}%
  \BibitemOpen
  \bibfield  {author} {\bibinfo {author} {\bibfnamefont {F.}~\bibnamefont
  {Zahid}}, \bibinfo {author} {\bibfnamefont {L.}~\bibnamefont {Liu}}, \bibinfo
  {author} {\bibfnamefont {Y.}~\bibnamefont {Zhu}}, \bibinfo {author}
  {\bibfnamefont {J.}~\bibnamefont {Wang}}, \ and\ \bibinfo {author}
  {\bibfnamefont {H.}~\bibnamefont {Guo}},\ }\href@noop {} {\bibfield
  {journal} {\bibinfo  {journal} {AIP Adv.}\ }\textbf {\bibinfo {volume} {3}},\
  \bibinfo {pages} {052111} (\bibinfo {year} {2013})}\BibitemShut {NoStop}%
\bibitem [{\citenamefont {Ruppert}\ \emph {et~al.}(2014)\citenamefont
  {Ruppert}, \citenamefont {Aslan},\ and\ \citenamefont
  {Heinz}}]{ruppert2014optical}%
  \BibitemOpen
  \bibfield  {author} {\bibinfo {author} {\bibfnamefont {C.}~\bibnamefont
  {Ruppert}}, \bibinfo {author} {\bibfnamefont {O.~B.}\ \bibnamefont {Aslan}},
  \ and\ \bibinfo {author} {\bibfnamefont {T.~F.}\ \bibnamefont {Heinz}},\
  }\href@noop {} {\bibfield  {journal} {\bibinfo  {journal} {Nano Lett.}\
  }\textbf {\bibinfo {volume} {14}},\ \bibinfo {pages} {6231} (\bibinfo {year}
  {2014})}\BibitemShut {NoStop}%
\bibitem [{\citenamefont {Rudenko}\ and\ \citenamefont
  {Katsnelson}(2014)}]{rudenko2014quasiparticle}%
  \BibitemOpen
  \bibfield  {author} {\bibinfo {author} {\bibfnamefont {A.~N.}\ \bibnamefont
  {Rudenko}}\ and\ \bibinfo {author} {\bibfnamefont {M.~I.}\ \bibnamefont
  {Katsnelson}},\ }\href@noop {} {\bibfield  {journal} {\bibinfo  {journal}
  {Phys. Rev. B}\ }\textbf {\bibinfo {volume} {89}},\ \bibinfo {pages}
  {201408(R)} (\bibinfo {year} {2014})}\BibitemShut {NoStop}%
\end{thebibliography}

%%Figures

\end{document}